\documentclass[twocolumn,twocolappendix]{aastex63}

\usepackage{amsmath, amsthm}
\usepackage{amsfonts,amssymb}
\usepackage{cancel}
\usepackage{subfigure}

 \received{February 2, 2022}
 \revised{March 4, 2022}
 \accepted{March 7, 2022}
\submitjournal{ApJL}
\shorttitle{Binary circularization periods}
\shortauthors{A. J. Barker}

\begin{document}

\title{Tidal dissipation due to inertial waves can explain the
circularization periods of solar-type binaries}

\correspondingauthor{Adrian Barker}
\email{A.J.Barker@leeds.ac.uk}

\author[0000-0003-4397-7332]{Adrian J. Barker}
\affiliation{Department of Applied Mathematics, School of Mathematics, University of Leeds, Leeds, LS2 9JT, UK}

\begin{abstract}
Tidal dissipation is responsible for circularizing the orbits and synchronizing the spins of solar-type close binary stars, but the mechanisms responsible are not fully understood. Previous work has indicated that significant enhancements to the theoretically-predicted tidal dissipation rates are required to explain the observed circularization periods ($P_\mathrm{circ}$) in various stellar populations, and their evolution with age. This was based partly on the common belief that the dominant mechanism of tidal dissipation in solar-type stars is turbulent viscosity acting on equilibrium tides in convective envelopes. In this paper we study tidal dissipation in both convection and radiation zones of rotating solar-type stars following their evolution. We study equilibrium tide dissipation, incorporating a frequency-dependent effective viscosity motivated by the latest hydrodynamical simulations, and inertial wave (dynamical tide) dissipation, adopting a frequency-averaged formalism that accounts for the realistic structure of the star. We demonstrate that the observed binary circularization periods can be explained by inertial wave (dynamical tide) dissipation in convective envelopes. This mechanism is particularly efficient during pre-main sequence phases, but it also operates on the main sequence if the spin is close to synchronism. The predicted $P_\mathrm{circ}$ due to this mechanism increases with main-sequence age in accord with observations. We also demonstrate that both equilibrium tide and internal gravity wave dissipation are unlikely to explain the observed $P_\mathrm{circ}$, even during the pre-main sequence, based on our best current understanding of these mechanisms. Finally, we advocate more realistic dynamical studies of stellar populations that employ tidal dissipation due to inertial waves.
\end{abstract} 


\section{Introduction}
\label{sec:intro}
Tidal interactions drive orbital and spin evolution in planetary systems and binary stars \citep[e.g.][]{Maezh2008,Zahn2008,Ogilvie2014}. Observations of various populations of stars with different ages provide strong evidence for efficient tidal dissipation in solar-type binary stars, both during the pre-main sequence (PMS) and also later on the main-sequence (MS) \citep[e.g.][]{Meibom2005,VanEylen2016,Triaud2017,Nine2020,Justesen2021}. The maximum orbital period out to which binary orbits are preferentially circular\footnote{See e.g.~\cite{Meibom2005} for a more detailed definition and \cite{Zanazzi2021} for a recent analysis and discussion.} is referred to as the circularization period, $P_\mathrm{circ}$. From analyzing samples of stars with different ages, $P_\mathrm{circ}$ has been found to increase with age on the main sequence. There is also evidence for spin synchronization in solar-type binaries \citep[e.g.][]{Meibom2006,Lurie2017} and circularization of evolved stars \citep[e.g.][]{VerbuntPhinney1995}.

Tidal flows in stars are often decomposed into two components \citep[e.g.][]{Zahn1977,Ogilvie2014}: a non-wavelike quasi-hydrostatic bulge and its associated flow that is referred to as the equilibrium tide, and a wavelike component that is referred to as the dynamical tide. The dynamical tide is thought to consist mainly of inertial waves in convection zones of rotating stars, and internal gravity waves in radiation zones.

Previous theoretical work has been unable to explain the observed $P_\mathrm{circ}$ and its variation with age. Early work by \cite{ZahnBouchet1989} (see also \citealt{Zahn2008}) suggested that dissipation of the equilibrium tide by the turbulent viscosity due to convection on the PMS could explain prior observations. However, their models assumed an optimistic turbulent viscosity acting on equilibrium tides\footnote{They also adopt an equilibrium tidal flow which is strictly invalid in large parts of convection zones \citep{Terquem1998,Ogilvie2014}, which likely over-estimates the dissipation by a factor of 2-3 \citep{B2020}, though this is unlikely to be a major problem.}, which is incompatible with the latest hydrodynamical simulations \citep[e.g.][]{OL2012,DBJ2020,DBJ2020a,VB2020,VB2020a}. In addition, it appears that another mechanism must operate later on the MS. Other works that accounted for a more realistic frequency-reduction of the turbulent viscosity have claimed that equilibrium tide dissipation is far too weak to explain the observed $P_\mathrm{circ}$ on the MS \citep{GO1997,Terquem1998}. This suggests that dissipation due to the dynamical tide must be responsible instead, in the absence of a more efficient mechanism to damp small-amplitude equilibrium tides\footnote{Such a mechanism has been proposed by \cite{TerquemMartin2021} but we believe they have likely significantly over-estimated the resulting dissipation \citep[see][]{BA2021}. This mechanism would be worth studying with further detailed simulations.}.

An obvious candidate is internal gravity wave dissipation in radiation zones. This appears to be more efficient than equilibrium tide dissipation for short orbital periods but is also unable to explain the observed $P_\mathrm{circ}$ \citep{GD1998,Terquem1998}. Since eccentricity tides in spin-synchronized (or possibly pseudo-synchronized; e.g.~\citealt{Hut1981}) binaries have forcing frequencies that are equal to (or similar to) the stellar spin frequency, this suggests that inertial waves (restored by Coriolis forces) could be important for tidal dissipation. Prior theoretical work has indicated that inertial waves can significantly enhance the dissipation of eccentricity tides, but this mechanism still appears to be insufficient to explain the observed $P_\mathrm{circ}$ in the MS models of \cite{OL2007}.

In this paper we build upon prior theoretical work by studying tidal dissipation in solar-type and low-mass binaries following their evolution from the PMS until they evolve off the MS. We demonstrate that inertial wave dissipation is very efficient in PMS phases, and it subsequently evolves with MS age in accord with observations. Inertial wave dissipation can therefore explain the observed $P_\mathrm{circ}$ and its variation with age (with equilibrium tides and gravity waves possibly contributing at the end of the MS).
We also demonstrate that both equilibrium tide dissipation and internal gravity wave dissipation are unlikely to account for the observed $P_\mathrm{circ}$, in agreement with the conclusions of prior work.

\section{Mechanisms of tidal dissipation}
\label{model}

We consider tides in both the convection and radiation zones of an approximately spherically-symmetric star with mass $M$ and radius $R$ in hydrostatic equilibrium, that rotates with angular velocity $\Omega=2\pi/P_\mathrm{rot}$ (where $P_\mathrm{rot}$ is the rotation period). We define the dynamical frequency $\omega_\mathrm{dyn}=\sqrt{GM/R^3}$ and timescale $P_\mathrm{dyn}=2\pi/\omega_\mathrm{dyn}$. We also define $\epsilon_\Omega^2 =P_\mathrm{dyn}^2/P_\mathrm{rot}^2$. The star is described by the standard equations of stellar structure, and we consider models with masses in the range $0.2-1.6 M_\odot$ computed using MESA \citep[e.g.][]{Paxton2011, Paxton2015,Paxton2019}, evolving them throughout the PMS phase until the end of the MS. Details of our calculations are relegated to Appendices~\ref{Appendix} and \ref{MESA}, and further details are reported in \cite{B2020}.

We compute the dissipation of equilibrium tides in convection zones by assuming a turbulent viscosity due to convection. This mechanism is believed to be strongly frequency-dependent, being inhibited in the regime of fast tides, when the tidal frequency $\omega$ exceeds the dominant convective turnover frequency $\omega_c$ \citep{Zahn1966,GN1977,Zahn1989}. We employ a frequency-dependent effective viscosity ($\nu_E$) that is compatible with the latest hydrodynamical simulations \citep{DBJ2020,DBJ2020a,VB2020,VB2020a}, which at high-frequency matches $\nu_E\propto (\omega/\omega_c)^{-2}$, the scaling law of \cite{GN1977}.

We calculate dynamical tide dissipation in radiation zones by assuming that internal gravity waves are launched from the radiative/convective interface and are then subsequently fully damped, following e.g.~\cite{GD1998} (who applied the ideas of e.g.~\citealt{Zahn1977} to solar-type stars). This provides a simple estimate of the dissipation due to these waves, which likely corresponds with the maximum dissipation in between resonances. Eccentricity tides in solar-type binaries are thought to be strongly nonlinear near the stellar centre \citep{GD1998,OL2007}, causing wave breaking \citep{BO2010,B2011}, following which these waves are expected to be efficiently damped.

Inertial wave dissipation is computed by employing a frequency-averaged formalism that fully accounts for the realistic internal structure of the star (following \citealt{Ogilvie2013} and \citealt{B2020}, which builds upon but is more realistic than the two-layer model considered by e.g.~\citealt{Mathis2015,Gallet2017}). This represents a crude measure of the tidal dissipation due to these waves that is primarily useful for population-wide studies such as this. This mechanism operates if the tidal frequency $\omega$ satisfies $|\omega|<2\Omega$, which is expected in spin-synchronized (or psuedo-synchronized) binaries because the relevant tidal frequencies for eccentricity tides are $\omega=\pm \Omega$ (or $|\omega|\sim\Omega$). We note that the dissipation due to inertial waves has been found to be strongly frequency-dependent in prior linear calculations \citep[e.g.][]{SavPap1997,OL2007,PapIv2010,Rieutord2010}. The frequency-averaged dissipation ignores this complicated behaviour but nevertheless provides a useful way to quantify the importance of inertial waves. It is also much simpler to calculate in stellar models.

\section{Tidal dissipation due to inertial waves}
\label{IWs}

\begin{figure}
  \begin{center} 
  \subfigure{\includegraphics[trim=3.5cm 0cm 5cm 0.5cm,clip=true,width=0.4\textwidth]{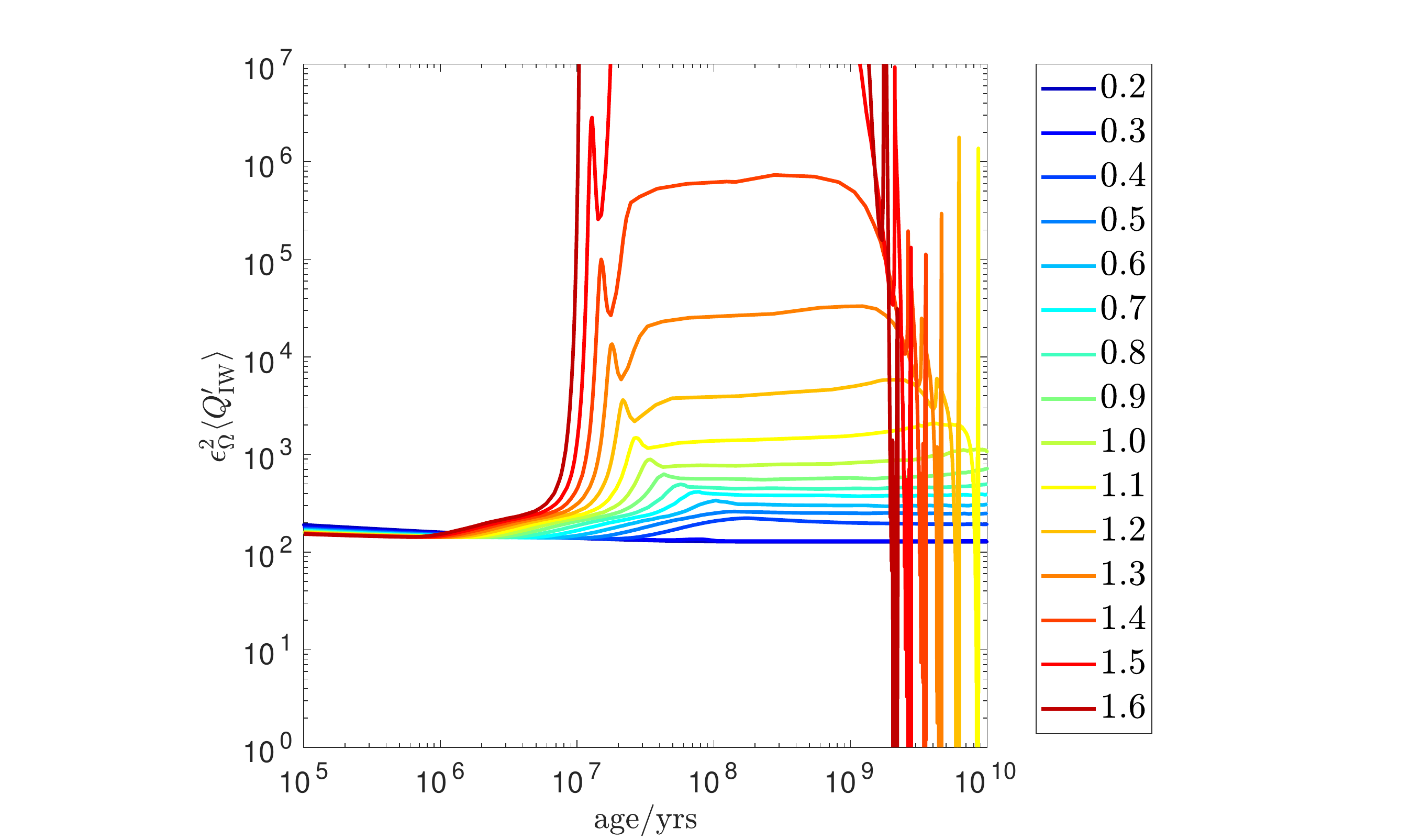}}
    \subfigure{\includegraphics[trim=3.5cm 0cm 5cm 0.5cm,clip=true,width=0.4\textwidth]{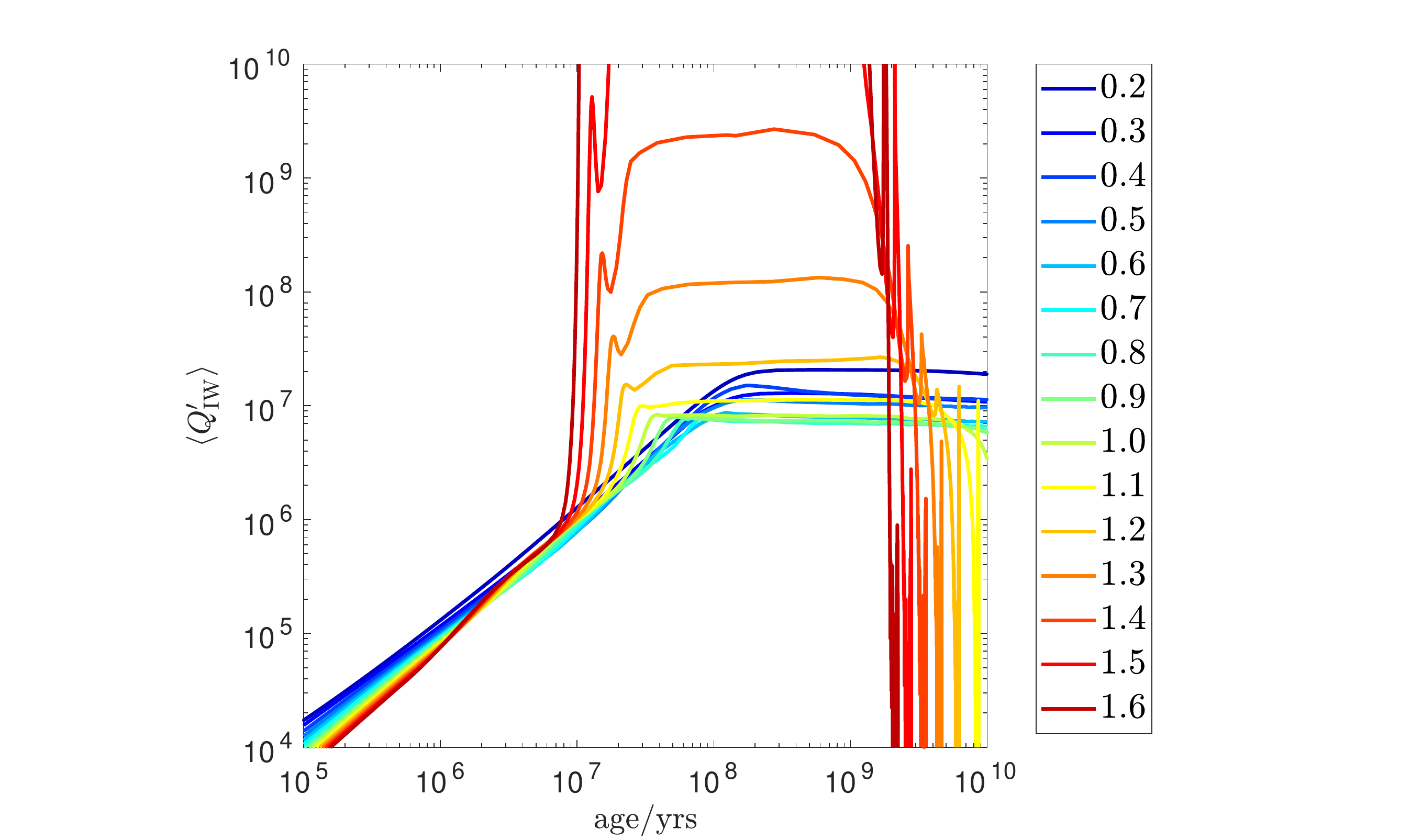}}
    \end{center}
  \caption{Top: tidal quality factor $\langle Q'_{\mathrm{IW}}\rangle$ multiplied by $\epsilon_\Omega^2$ as a function of age, due to (frequency-averaged) dissipation of inertial waves in the convective envelopes of various stars with the masses (in solar masses) indicated in the legend. Bottom: same, except that $P_\mathrm{rot}=10$ d following the evolution of each star. The most efficient dissipation is found during both the PMS and as the star evolves off the MS.}
  \label{QpIW}
\end{figure}

\begin{figure}
  \begin{center} 
  \subfigure{\includegraphics[trim=3.5cm 0cm 5cm 1cm,clip=true,width=0.4\textwidth]{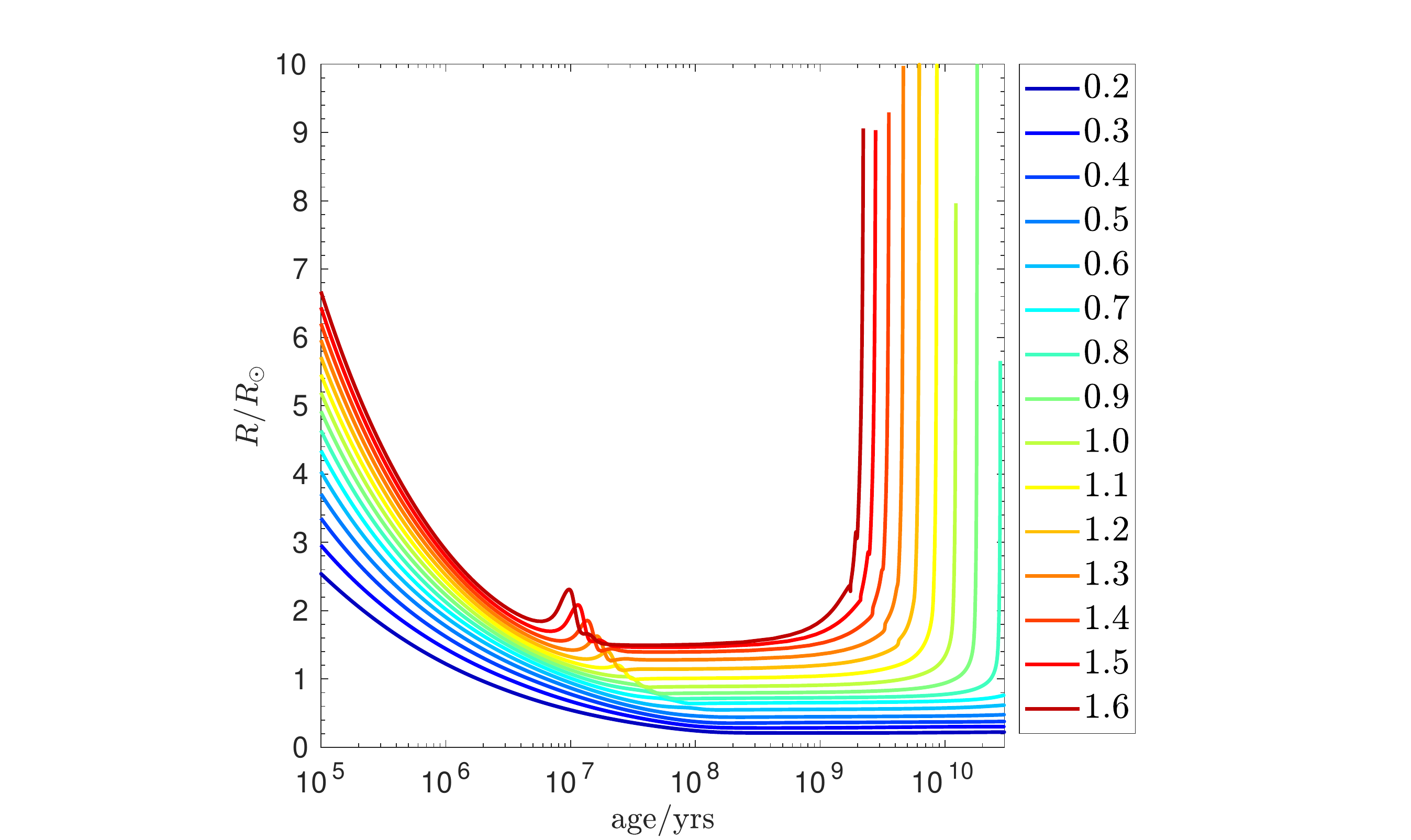}}
    \end{center}
  \caption{Evolution of the stellar radius $R$ (normalized by the solar radius $R_\odot$) with age ($t$) for the stars in our mass range. The radius is largest during the PMS ($t\lesssim 10^7$ yr) and in later phases as the stars evolve off the MS ($t\gtrsim O(10^9)$ yr).}
  \label{RoverRodot}
\end{figure}

We first introduce our results for the frequency-averaged dissipation due to inertial waves by plotting a modified tidal quality factor\footnote{This is related to the usual $Q$ by $Q'=3Q/(2k_2)$, where $k_2$ is the 2nd order potential Love number.} for this component $Q'\equiv\langle Q'_{\mathrm{IW}}\rangle$, which is an inverse measure of the dissipation, as a function of stellar age in Fig.~\ref{QpIW}. We have shown results for stars with masses in the range $0.2-1.6 M_\odot$. The top panel indicates $\epsilon_\Omega^2 \langle Q'_\mathrm{IW} \rangle$. If the rotation period $P_\mathrm{rot}$ of a star is known (along with its mass and radius), then this figure allows the typical level of tidal dissipation due to inertial waves to be computed. We complement this in the bottom panel, where we have shown $\langle Q'_\mathrm{IW} \rangle$ by setting $P_\mathrm{rot}=10$ d for all stars.

Fig.~\ref{QpIW} shows that inertial wave dissipation is very efficient during PMS phases and evolves to become less efficient on the MS (until the latest stages). This conclusion will be further strengthened by considering that young stars rotate more rapidly. Cooler stars are generally more dissipative than hotter stars, in that they have smaller $\langle Q'_\mathrm{IW} \rangle$ on the MS, with F-stars (with masses $M>1.3 M_\odot$) being the least dissipative in this mass range. For stars with masses in the range $0.2-1.2 M_\odot$, we typically find $\langle Q'_\mathrm{IW}\rangle \approx 10^7 (P_\mathrm{rot}/10\, \mathrm{d})^2$.

Since tidal timescales strongly depend on the stellar radius, it is essential to explore how this varies with age. In Fig.~\ref{RoverRodot}, we plot the evolution of the stellar radius with age, which highlights that the PMS and the later evolutionary phases at the end of the MS are those in which these stars have the largest radii. Together with the results of Fig.~\ref{QpIW}, we might therefore expect the PMS and the later evolutionary phases at the end of the MS to be the most important for tidal dissipation.

\section{Circularization periods}
\label{CircTheory}

\begin{figure}
  \begin{center}
    \subfigure[Inertial waves]{\includegraphics[trim=4cm 0cm 5cm 1cm,clip=true,width=0.4\textwidth]{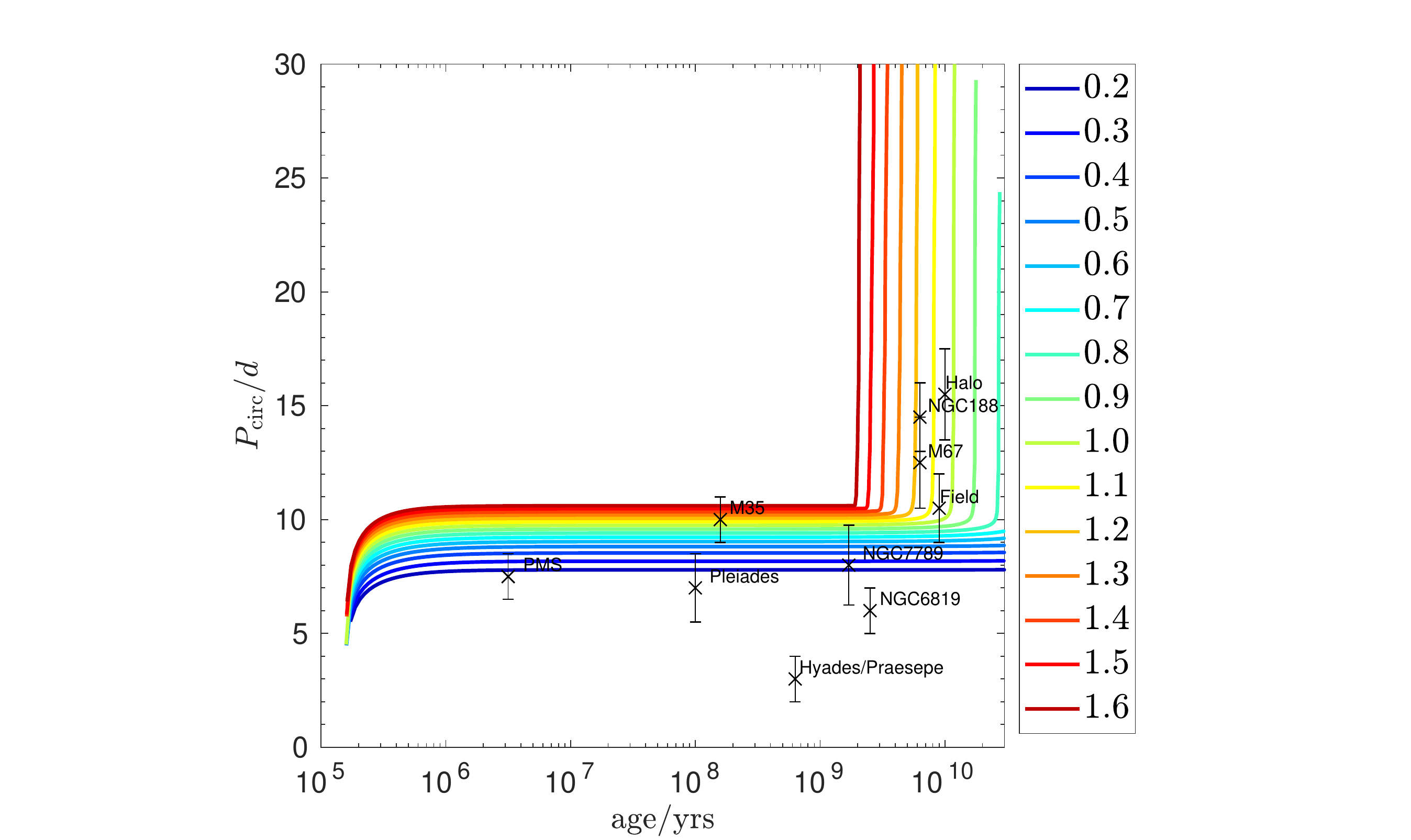}}
    \subfigure[Internal gravity waves]{\includegraphics[trim=4cm 0cm 5cm 1cm,clip=true,width=0.4\textwidth]{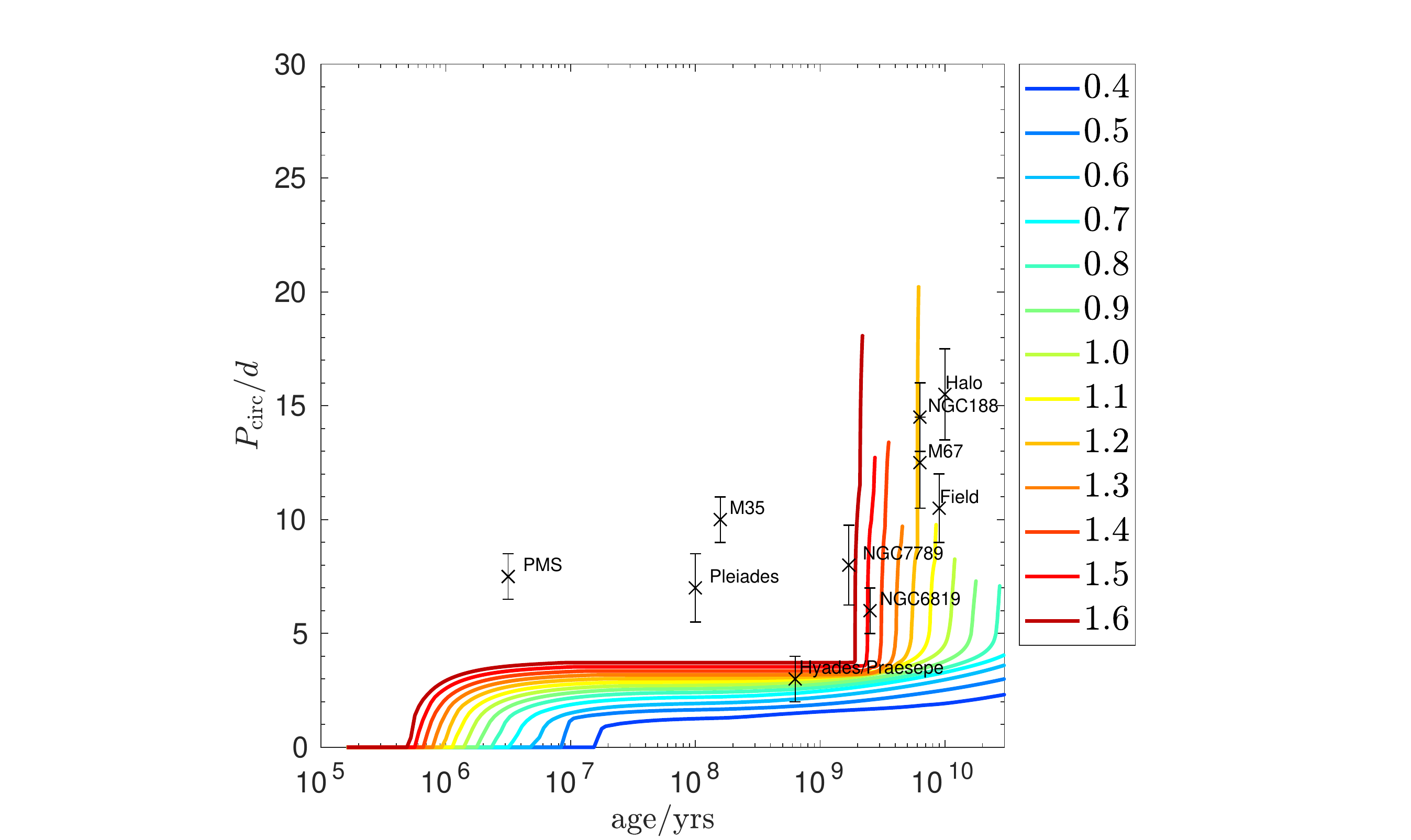}}
    \subfigure[Equilibrium tides]{\includegraphics[trim=4cm 0cm 5cm 1cm,clip=true,width=0.4\textwidth]{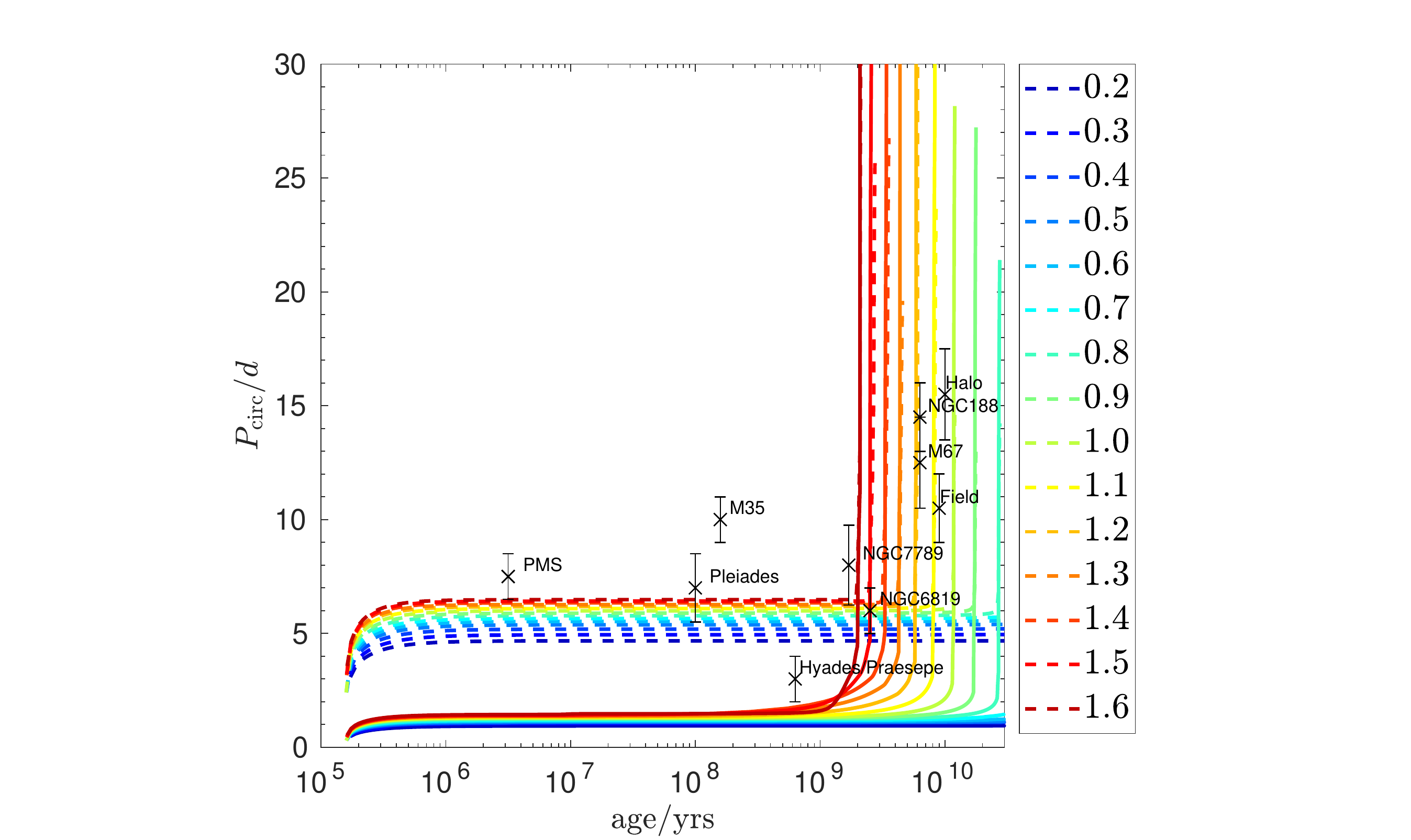}}
    \put(-170,65){$\nu_E=u_cl_c/3$}
    \put(-170,73){(unphysical)}
    \put(-170,30){$\nu_E=\nu_{FIT}$}
    \end{center}
  \caption{$P_\mathrm{circ}$ as a function of age (assuming $P_\mathrm{rot}=P_\mathrm{orb}$), due to dissipation of inertial waves (top), internal gravity waves (middle) and equilibrium tides (bottom; where $\nu_E$ is equal to the high frequency portion of $\nu_{FIT}$ in solid and $\nu_E=u_cl_c/3$ in dashed -- note: the latter is unphysical), computed for a range of stellar masses (legend, in solar masses). We over-plot the observations compiled by \cite{Nine2020} as black symbols with labels. Inertial wave dissipation can explain the observed binary circularization periods.}
  \label{Pcirccomp}
\end{figure}

We now apply the results in Figs.~\ref{QpIW} and \ref{RoverRodot} to calculate theoretically the circularization periods of binary stars due to inertial waves. We then calculate $P_\mathrm{circ}$ due to equilibrium tides and internal gravity waves for comparison. For a conservative estimate, we consider dissipation of the $l=m=2$ (spherical harmonic degree $l$ and azimuthal wavenumber $m$) tide in only the primary star of mass $M$ that has a companion of mass $M_2=M_\odot$ (for all primary stars), and we assume spin-orbit synchronization ($P_\mathrm{rot}=P_\mathrm{orb}$, the orbital period; which we will later justify for most evolutionary stages). For small $e$, the eccentricity evolves according to\footnote{This can be obtained from Eq.~5 in \cite{OL2007} (or Eq.~1 in \citealt{ZahnBouchet1989}) by neglecting the component with $m=0$ and assuming $Q'$ is the same for frequencies $\pm \Omega$.}
\begin{eqnarray}
\frac{\mathrm{d}\ln e}{\mathrm{d} t} = -\frac{225\pi}{8}\frac{1}{Q'}\left(\frac{M_2}{M}\right)\left(\frac{M}{M+M_2}\right)^{\frac{5}{3}}\frac{P_{\mathrm{dyn}}^{\frac{10}{3}}}{P_\mathrm{orb}^{\frac{13}{3}}},
\end{eqnarray}
where $Q'$ is an appropriate tidal quality factor. This equation can be integrated from a (uncertain) starting age $t_0$ to a final age $t$, following stellar evolution, to give
\begin{eqnarray}
\label{dlnedt}
\hspace{-0.4cm}
\Delta \ln e = \frac{225\pi}{8}\left(\frac{M_2}{M}\right)\left(\frac{M}{M+M_2}\right)^{\frac{5}{3}}\frac{1}{P_\mathrm{orb}^{\frac{13}{3}}} \int_{t_0}^{t}\frac{P_{\mathrm{dyn}}^{\frac{10}{3}}}{Q'}\,\mathrm{d} t ,
\end{eqnarray}
where $\Delta \ln e=\ln e(t_0)-\ln e(t)$.
For inertial wave dissipation, assuming spin-orbit synchronization, we set $Q' = \left[\epsilon_\Omega^2\langle Q'_\mathrm{IW}\rangle\right] P_\mathrm{orb}^2/P_\mathrm{dyn}^2$, such that
\begin{eqnarray}
\nonumber
\Delta \ln e = \frac{225\pi}{8}\left(\frac{M_2}{M}\right)\left(\frac{M}{M+M_2}\right)^{\frac{5}{3}}\frac{1}{P_\mathrm{orb}^{\frac{19}{3}}} \int_{t_0}^{t}\frac{P_{\mathrm{dyn}}^{\frac{16}{3}}}{\epsilon_\Omega^2\langle Q'_\mathrm{IW}\rangle}\,\mathrm{d} t.
\end{eqnarray}
Note that for low-mass stars in which $\epsilon^2_\Omega\langle Q'_\mathrm{IW}\rangle$ is approximately constant as they evolve from the PMS to the end of the MS (see Fig.~\ref{QpIW}), $\langle Q'_\mathrm{IW}\rangle\propto 1/R^3$, indicating that $\Delta \ln e \propto R^{11}$, an extremely strong function of $R$. This indicates that even though the PMS phase is very short, since the star then has much larger $R$ it can dominate the circularization \citep[as also hypothesized for equilibrium tides by][]{ZahnBouchet1989}.

For a conservative estimate here we assume an initial eccentricity of 1 (albeit strictly invalidating the assumption of small $e$) at $t_0=0.15\,\mathrm{Myr}$ and a final eccentricity of 0.01 at each age $t>t_0$, so that $\Delta \ln e\approx 4.6$. This equation is rearranged to obtain the critical orbital period $P_\mathrm{circ}\equiv P_\mathrm{orb}$ out to which circularization is expected, assuming the required value of $\Delta \ln e$, and computing the integral following the evolution of the star. The integrand is computed using hundreds of snapshots following stellar evolution, which are interpolated to a finer grid in time to calculate the integral numerically.

Our results for $P_\mathrm{circ}$ as a function of age are shown in the top panel of Fig.~\ref{Pcirccomp} for stars with masses $0.2-1.6 M_\odot$. This demonstrates that efficient dissipation of inertial waves on the PMS can circularize orbits out to 10 d prior to a few Myrs. There is then negligible evolution of $P_\mathrm{circ}$ on the MS until a few Gyr (depending on $M$), after which $P_\mathrm{circ}$ increases again as the star evolves towards the end of its life on the MS. These theoretical predictions are in accord with the observations compiled by \cite{Nine2020}, which are indicated as the black symbols with labels on this figure. Fig.~\ref{Pcirccomp} therefore demonstrates that inertial wave dissipation is able to explain the observed circularization periods of close binary stars. We do not need to invoke any artificial enhancement of tidal dissipation to explain these observations. Note that $t_0$ was chosen to explain M35 since this requires the most efficient dissipation; choosing e.g.~$t_0\approx 0.3$ Myr instead allows the PMS and Pleiades to be explained neatly with the correct stellar masses without affecting later MS stages.

In the middle panel of Fig.~\ref{Pcirccomp}, we show the corresponding prediction for $P_\mathrm{circ}$ due to internal gravity wave dissipation. This is obtained by solving Eq.~\ref{dlnedt} after setting $Q'=Q'_\mathrm{IGW}$, as defined by Eq.~\ref{QIGW}, which leads to $\Delta \ln e \propto P_\mathrm{orb}^{-7}$. This mechanism is insufficient to explain the observed $P_\mathrm{circ}$ until later ages on the MS, when it also predicts an increase in $P_\mathrm{circ}$ with age, though only for the most massive stars with radiative cores.

Finally, we show the same result for equilibrium tide dissipation in the bottom panel of Fig.~\ref{Pcirccomp}, which is obtained by solving Eq.~\ref{dlnedt} with $Q'=Q'_\mathrm{eq}$, as defined by Eq.~\ref{Qeq}, which gives $\Delta \ln e \propto P_\mathrm{orb}^{-10/3}$ in the high-frequency regime for $\nu_E=\nu_{FIT}$ in Eq.~\ref{nuEFIT1} (the same conclusion is reached using $\nu_E$ based on \citealt{GN1977}). This is shown using solid lines, and a similar prediction that instead uses the (unphysical) conventionally-adopted mixing-length expectation (with no frequency-reduction) $\nu_E=u_cl_c/3$, for which $\Delta \ln e \propto P_\mathrm{orb}^{-16/3}$, is shown using dashed lines. This mechanism is clearly inefficient, and is unable to circularize orbits outside 2 days prior to Gyr ages due to the reduction in the turbulent viscosity for high frequencies. This disagrees with the conclusions of \cite{ZahnBouchet1989} because they adopted a less drastic frequency-reduction for $\nu_E$, which is incompatible with the latest (albeit idealized) simulations \citep{OL2012,DBJ2020,DBJ2020a,VB2020,VB2020a}. On the other hand, we also show that even if there is no reduction in $\nu_E$ for high frequencies, while this mechanism is then much more efficient, it is still unable to explain the observations even during the PMS. This mechanism becomes more efficient however as the star evolves towards the end of the MS, when it can provide a comparable contribution to inertial waves (for either choice of $\nu_E$).

Our estimates for $P_\mathrm{circ}$ here are probably overly conservative estimates for several reasons: 1) we have considered tides in only the primary star (considering tides in both stars would raise $P_\mathrm{circ}$ due to inertial waves by approximately $2^{3/19}\approx 12\%$), 2) we assume a large $\Delta \ln e$, 3) we start at $t_0=0.15 \mathrm{Myr}$, meaning that we ignore earlier (short) PMS phases when the star had an even larger radius, 4) we ignore tidal components other than $l=m=2$, 5) we assume spin-orbit synchronization, which may overestimate the rotation periods for young stars. We have found that using an earlier $t_0<0.1$ Myr can lead to \textit{significantly larger} $P_\mathrm{circ}$ for inertial wave dissipation i.e.~this mechanism is then \textit{too efficient} to fit the observations. Our choice of $t_0=0.15$ Myr here was found to give a good fit with the observations for the stellar models considered, as we have shown in Fig.~\ref{Pcirccomp}, though values $t_0\lesssim 0.8$ Myr also work well (except for M35). On the other hand, the frequency-averaged measure may over or under-estimate the dissipation of inertial waves at a given tidal frequency by an uncertain factor that is hard to quantify. Further work is required to explore this matter by directly solving the linear tidal response for eccentricity tides in a wide range of stellar models.

\begin{figure}
  \begin{center}
  \subfigure{\includegraphics[trim=4cm 0cm 5cm 1cm,clip=true,width=0.4\textwidth]{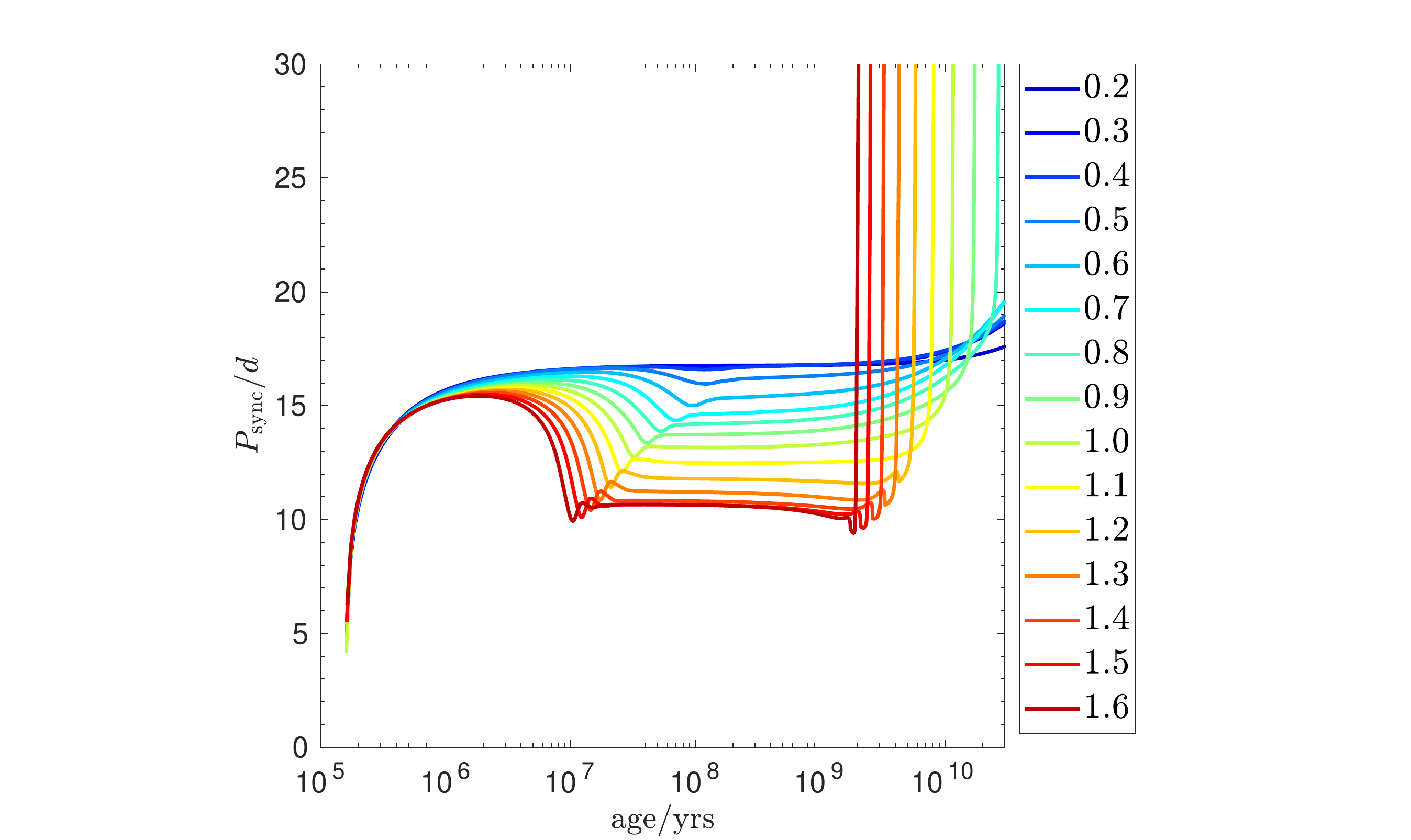}}
    \end{center}
  \caption{Synchronization period $P_\mathrm{sync}$, assuming an initial rotation period $P_\mathrm{rot}=5$ d. Since $P_\mathrm{sync}(t)>P_\mathrm{circ}(t)$ due to inertial waves, this indicates that spin-orbit synchronization is a reasonable assumption for the purposes of Fig.~\ref{Pcirccomp}.}
  \label{Psynccomp}
\end{figure}

\begin{figure}
  \begin{center}
  \subfigure{\includegraphics[trim=2.5cm 0cm 3.75cm 1cm,clip=true,width=0.4\textwidth]{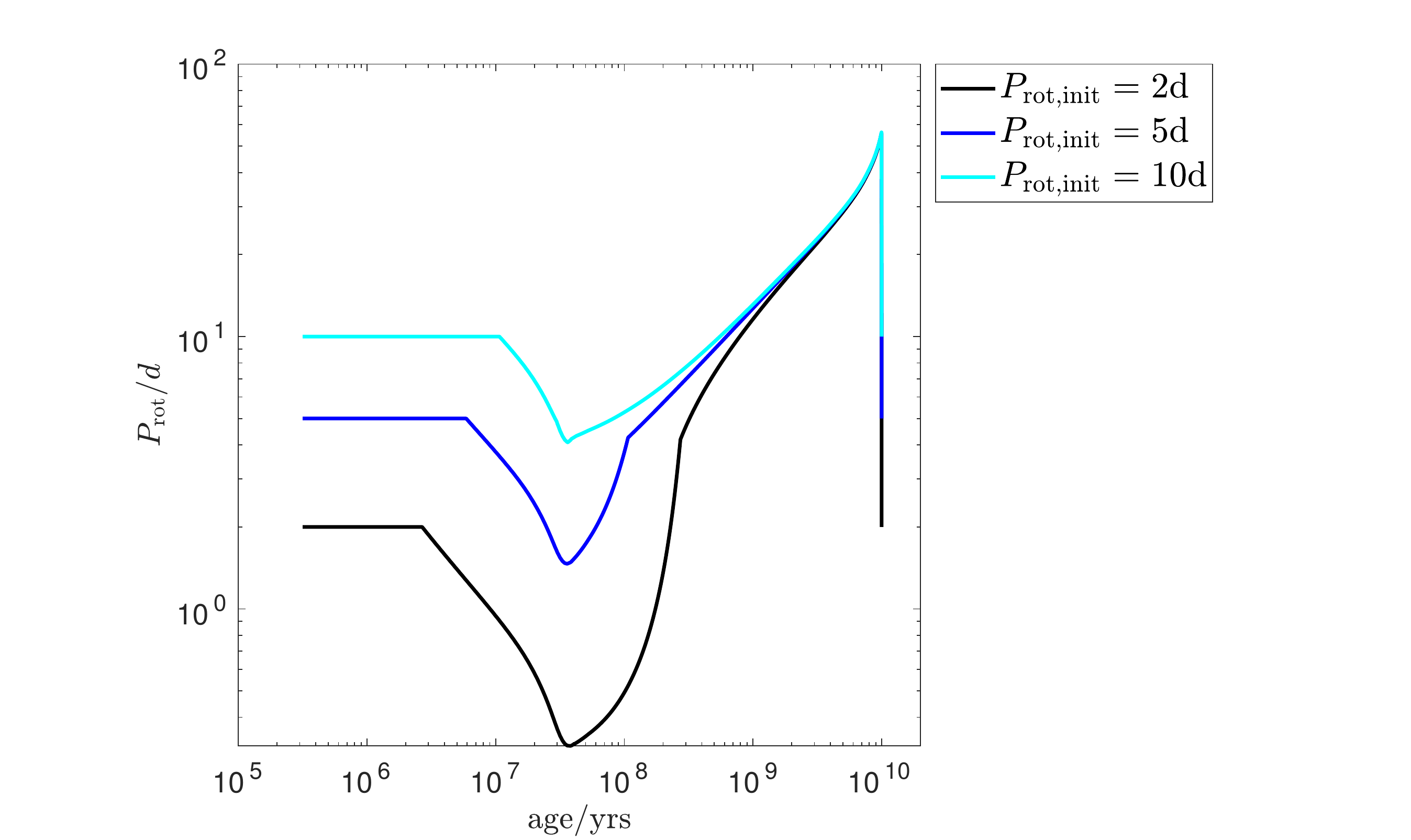}}
  \subfigure{\includegraphics[trim=2.75cm 0cm 3.75cm 1cm,clip=true,width=0.4\textwidth]{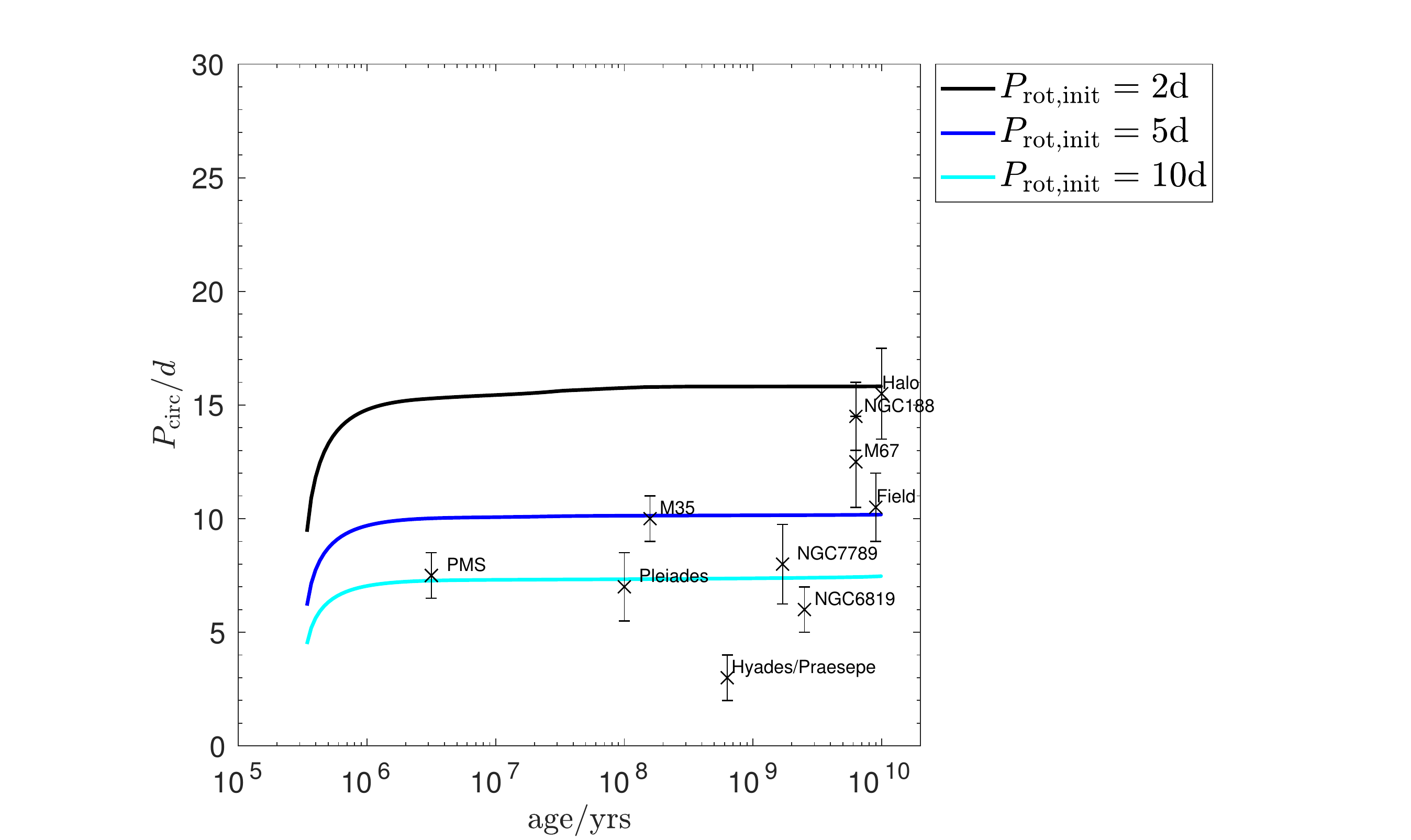}}
    \end{center}
  \caption{Top: $P_{\mathrm{rot}}$ as a function of age for a solar-mass star with three initial rotation periods computed with star-disc locking and magnetic braking but ignoring tides. Bottom: $P_\mathrm{circ}$ from solving Eq.~\ref{dlnedt} for inertial wave dissipation using the above $P_\mathrm{rot}$ as a function of age. This can be compared with Fig.~\ref{Pcirccomp}, and indicates that PMS circularization is efficient without assuming $P_\mathrm{orb}=P_\mathrm{rot}$.}
  \label{Pcircrot}
\end{figure}

We can crudely demonstrate that our assumption of spin-orbit synchronization above is not unreasonable by solving\footnote{The dimensionless squared radius of gyration is 
\begin{eqnarray}
r_g^2 = \frac{8\pi}{GM R^2}\int_0^R\rho r^4\mathrm{d}r.
\end{eqnarray}}
\begin{eqnarray}
\nonumber
\Delta \ln P_\mathrm{rot} = \frac{9\pi r_g^2}{2}\left(\frac{M_2}{M+M_2}\right)^{2}\frac{1}{P_{\mathrm{rot}}P_{\mathrm{orb}}^4} \int_{t_0}^{t}\frac{P_{\mathrm{dyn}}^{4}}{\epsilon_\Omega^2\langle Q'_\mathrm{IW}\rangle}\,\mathrm{d} t.
\end{eqnarray}
We assume $P_\mathrm{rot}=5$ d and we define $P_\mathrm{sync}\equiv P_\mathrm{orb}$ by assuming the logarithmic change in the stellar rotation period $\Delta\ln P_\mathrm{rot}=1$. The results are shown in Fig.~\ref{Psynccomp}, which indicates the typical orbital period out to which tidal evolution of stellar spin is efficient. Since $P_\mathrm{circ}$ is generally shorter than $P_\mathrm{sync}$, it is reasonable to assume spin-orbit synchronization for most ages in Fig.~\ref{Pcirccomp}. 

The efficiency of inertial wave PMS circularization is further evidenced by Fig.~\ref{Pcircrot}. Here we model the rotational evolution of our solar-mass star (ignoring tides) through the PMS to the end of the MS by following the approaches of e.g.~\cite{Amard2019,Lazovik2021}, including star-disc locking and magnetic braking \citep[following e.g.][]{Matt2015}. The top panel shows the evolution of $P_\mathrm{rot}$ as a function of age for three initial rotation periods, and the bottom panel shows the corresponding evolution of $P_\mathrm{circ}$ due to inertial waves by solving Eq.~\ref{dlnedt} (using $t_0=0.32$ Myr) for these $P_\mathrm{rot}(t)$ (instead of assuming $P_\mathrm{rot}=P_\mathrm{orb}$). This can be compared with Fig.~\ref{Pcirccomp}, and indicates that efficient inertial wave PMS circularization is predicted without assuming spin-orbit synchronization. A more realistic calculation should really account for the tidal evolution of $\Omega$ along with $e$, accounting for multiple tidal components. For example, eccentricity can be briefly excited by tides when the star rotates sufficiently rapidly relative to the orbit \citep[e.g.][]{ZahnBouchet1989}, and stellar magnetic braking rates are probably modified by tidal evolution of the stellar rotation, which may both modify slightly our conclusions for $P_\mathrm{circ}$. However, such a more detailed study is outside the scope of this letter.

Our results are in conflict with \cite{ZahnBouchet1989}, who claim that ``there is little doubt that the circularization in these stars is due to the action of the equilibrium tide early on the PMS". We do agree that the PMS phase is probably the most important for tidal circularization, mainly because the stars then have much larger radii than on the MS \citep[see also][]{Gallet2017} -- at least based on standard models for PMS evolution \citep[e.g.][]{StahlerPalla2005}; the picture may be different in ``cold accretion" models where the stars do not attain as large radii \citep[e.g.][]{Kunitomo2017}, such that understanding tidal evolution better may even be able to constrain PMS evolution. However, while significant uncertainties remain, based on our current understanding of tidal dissipation, inertial waves are probably much more important in this problem, and they can readily explain the observed $P_\mathrm{circ}$ and its evolution with age.

It is interesting to briefly discuss our results in light of the recent analysis of observational data for solar-mass binaries by \cite{Zanazzi2021}, who argues against the validity of the longer $P_\mathrm{circ}$ from observations that we have reproduced in Fig.~\ref{Pcirccomp}. They found evidence for $P_\mathrm{circ}\approx 3$ days for eccentric binaries, approximately independent of MS age, but with a population of circular binaries out to 10 days that they suggest could bias previous studies with smaller sample sizes \citep[e.g.][]{Meibom2005}, and perhaps produce their larger inferred values of $P_\mathrm{circ}$. We have shown here that inertial waves are efficient enough to be able to explain the $P_\mathrm{circ}$ previously reported, as we show in Fig.~\ref{Pcirccomp}. The two populations identified by \cite{Zanazzi2021} could in principle be produced by a distribution in initial stellar rotation periods, which we have shown can potentially lead to very different $P_\mathrm{circ}$ in Fig.~\ref{Pcircrot}. Since stars that have undergone efficient circularization would generally have synchronized even more rapidly (Fig.~\ref{Psynccomp}), analyzing stellar rotation periods provides a possible way to observationally infer whether inertial waves have been responsible, or whether the nearly circular orbits are produced without strong stellar tidal dissipation e.g.~by eccentricity damping during gas-disc migration\footnote{For comparison with Fig.~\ref{Pcirccomp}, if we instead set $t_0=10$ Myr (after dispersal of gas discs), inertial waves circularize orbits on the PMS out to 3 days, with subsequent gradual evolution of $P_\mathrm{circ}$ on the MS.}, perhaps with eccentric binaries produced by dynamical interactions. Our results for solar-mass stars are consistent with \cite{Zanazzi2021}'s observation that most tidal evolution occurs during the PMS, and that there is negligible evolution on the MS until the latest evolutionary stages $\gtrsim 5$ Gyr.

\section{Conclusions} \label{Conclusions}

We have demonstrated that the circularization periods of solar-type binary stars \citep[e.g][]{Meibom2005,Triaud2017,Nine2020} can be explained by tidal dissipation due to inertial waves in convection zones. This mechanism is very efficient during the PMS; it also predicts an increase in $P_\mathrm{circ}$ with age for spin-synchronized stars on the MS in accord with observations. 

We have shown that this component of the dynamical tide is much more dissipative than convective damping of equilibrium tides, both during the PMS and until the latest evolutionary stages on the MS. This is primarily because the latter mechanism is strongly inhibited by the frequency-reduction of the turbulent viscosity \citep{GN1977}. The latter result agrees with \cite{GO1997} and \cite{Terquem1998} but disagrees with \cite{ZahnBouchet1989}, who adopted a much weaker frequency-reduction that is inconsistent with the latest simulations \citep{OL2012,DBJ2020,DBJ2020a,VB2020,VB2020a}. We also show that internal gravity wave dissipation is unlikely to explain the observed circularization periods, in agreement with \cite{GD1998}.

Previous theoretical work studying tidal dissipation in rotating solar-type stars suggested that inertial wave dissipation could be the dominant mechanism for binary circularization \citep{OL2007}. However, this mechanism was previously thought to be too weak to account for the observed circularization periods of solar-type binary stars on the MS \citep[see also][]{B2020}. Our new conclusion here has arisen following inertial wave dissipation along the evolution of solar-type stars. We have demonstrated that inertial wave dissipation is very efficient during PMS phases, and that it evolves with MS age, such that it can readily produce the observed circularization periods of solar-type binaries. Inertial waves have the additional benefit that they do not pose a problem for the survival of the shortest-period hot Jupiters; indeed they are not linearly excited for aligned orbits unless $P_\mathrm{rot}\leq 2P_\mathrm{orb}$,
and so cannot drive orbital decay around slowly-rotating stars \citep[e.g.][]{OL2007}.

Our idealized proof-of-concept calculations have focused on tidal evolution of binary eccentricities by modelling the dissipation due to inertial waves using a highly simplified frequency-averaged formalism that properly accounts for the realistic internal structure of the star (following \citealt{Ogilvie2013}). Future work should study the impact of this mechanism of tidal dissipation in more sophisticated calculations that also study the dynamical evolution of stellar populations \citep[e.g.][]{Geller2013} and simultaneously model tidal dissipation with models for the evolution of stellar rotation from the PMS until the end of the MS \citep[e.g.][]{Gallet2017}. It would also be worthwhile to explore the excitation of inertial waves in calculations that directly compute the frequency-dependent linear tidal response in a range of stellar models, and to explore further the possibility of resonance locking\footnote{Resonance locking of gravity modes is however unlikely to explain circularization of solar-type binaries due to wave breaking in radiative cores, which can prevent g-mode resonances.} following stellar evolution \citep[e.g.][]{WitteSavonije2002,ZanazziWu2021}.

\acknowledgments
This work was funded by STFC Grants ST/R00059X/1 and ST/S000275/1. We would like to thank the reviewer, Robert Mathieu and Gordon Ogilvie for their very helpful comments and suggestions.

\bibliographystyle{aasjournal}
\bibliography{tid}{}
\appendix
\section{Method} \label{Appendix}

We consider a single $l=m=2$ tidal potential component
\begin{eqnarray}
\Psi(\boldsymbol{r},t)&=& \mathrm{Re}\left[\Psi_l (r) Y_l^m(\theta,\phi) \mathrm{e}^{-\mathrm{i}\omega t} \right],
\end{eqnarray}
inside the primary star of mass $M$, due to another star of mass $M_2$ in a close binary system, where we define $\Psi_l=A r^l$, where $A$ is the tidal amplitude and $(r,\theta,\phi)$ are the usual spherical coordinates centred on the primary \citep[e.g.][]{Ogilvie2014}. The linearized Eulerian gravitational potential response is 
\begin{eqnarray}
\Phi'(\boldsymbol{r},t) = \mathrm{Re}\left[\Phi'_l(r) Y_l^m(\theta,\phi) \mathrm{e}^{-\mathrm{i}\omega t} \right],
\end{eqnarray}
with a similar expansion for other variables. 

In a convective region with efficient convection such that it is adiabatically stratified, the low-frequency equilibrium tide is irrotational \citep{Terquem1998,GD1998},
\begin{eqnarray}
\boldsymbol{\xi}_\mathrm{nw}=\nabla X,
\end{eqnarray}
and its displacement satisfies, after expanding $X$ in terms of spherical harmonics \citep{Ogilvie2013},
\begin{eqnarray}
\label{Xeqn}
\hspace{-0.5cm}
\frac{1}{r^2}\frac{\mathrm{d}}{\mathrm{d}r}\left(r^2\rho \frac{\mathrm{d} X_l }{\mathrm{d}r}\right)-\frac{l(l+1)}{r^2}\rho X_l=-\rho\frac{\mathrm{d}\rho}{\mathrm{d}p}(\Phi'_l+\Psi_l),
\end{eqnarray}
where $\rho(r)$ is the density and $p(r)$ is the pressure. The boundary conditions on a convective core are
\begin{eqnarray}
\xi_{\mathrm{nw},r}=\frac{\mathrm{d}X_l}{\mathrm{d}r}=0 \quad \text{at} \quad r=0,
\end{eqnarray}
and for all other boundaries of any convection zone,
\begin{eqnarray}
\xi_{\mathrm{nw},r}=\frac{\mathrm{d}X_l}{\mathrm{d}r}=-\frac{\Phi'+\Psi}{g},
\end{eqnarray}
which also applies at $r=R$ for a convective envelope, where $g(r)$ is the gravitational acceleration. The perturbation to the gravitational potential satisfies 
\begin{eqnarray}
\label{PhiEqn}
\nonumber
&& \frac{1}{r^2}\frac{\mathrm{d}}{\mathrm{d}r}\left(r^2 \frac{\mathrm{d} \Phi'_l }{\mathrm{d}r}\right)-\frac{l(l+1)}{r^2}\Phi'_l \\
&& \hspace{1.25in}
+4\pi G\frac{\mathrm{d}\rho}{\mathrm{d}p} \rho(\Phi'_l+\Psi_l)=0,
\end{eqnarray}
where $\Phi'_l$ must satisfy the boundary conditions
\begin{eqnarray}
&&\frac{\mathrm{d}\ln \Phi'_l}{\mathrm{d} \ln r} = l \quad\text{at} \quad r=0, \\
&&\frac{\mathrm{d}\ln \Phi'_l}{\mathrm{d} \ln r} = -(l+1) \quad\text{at} \quad r=R.
\end{eqnarray}

The equilibrium tide defined here differs from the conventional equilibrium tide of e.g.~\cite{Zahn1989}. The latter is strictly invalid in large parts of convection zones, and we find that it also typically over-predicts the dissipation by a factor of 2-3 \citep{B2020}. Our main conclusions would however be unchanged if we were to adopt the conventional equilibrium tide here instead.

Following \cite{Ogilvie2013}, the pressure perturbation of inertial waves (proportional to $W_l Y_l^m $, but obtained using an impulsive formalism) satisfies
\begin{eqnarray}
\label{WL}
\hspace{-0.5cm}
\frac{1}{r^2}\frac{\mathrm{d}}{\mathrm{d}r}\left(r^2\rho \frac{\mathrm{d} W_l }{\mathrm{d}r}\right)-\frac{l(l+1)}{r^2}\rho W_l=\frac{2\mathrm{i}m\Omega}{r}\frac{\mathrm{d}\rho}{\mathrm{d} r}X_l,
\end{eqnarray}
subject to the boundary conditions of vanishing radial velocity at the boundaries of each convection zone, i.e.
\begin{eqnarray}
\frac{\mathrm{d} W_l }{\mathrm{d}r}=\frac{2\mathrm{i}m\Omega X_l}{r}.
\end{eqnarray}
The components of the flow are
\begin{eqnarray}
a_l &=& \frac{2\mathrm{i}m\Omega}{r}X_l-\frac{\mathrm{d}W_l}{\mathrm{d}r}, \\
b_l &=& \frac{2\mathrm{i}m\Omega}{l(l+1) r^2}\left(r \frac{\mathrm{d}X_l}{\mathrm{d}r}+X_l\right)-\frac{W_l}{r^2}, \\
c_{l-1} &=& \frac{2\Omega q_l}{r^2}\left(r \frac{\mathrm{d}X_l}{\mathrm{d}r}+(l+1) X_l\right), \\
c_{l+1} &=& -\frac{2\Omega q_{l+1}}{r^2}\left(r \frac{\mathrm{d}X_l}{\mathrm{d}r}-l X_l\right)
\end{eqnarray}
where 
\begin{eqnarray}
q_l = \frac{1}{l}\left(\frac{l^2-m^2}{4l^2-1}\right)^{\frac{1}{2}}.
\end{eqnarray}
The associated tidal quality factor representing the frequency-averaged dissipation of inertial waves is then
\begin{eqnarray}
\label{QIW}
\hspace{-0.6cm}
\frac{1}{\langle Q'_{\mathrm{IW}}\rangle} = \frac{32\pi^2G}{3(2l+1)R^{2l+1}|A|^2}(E_l+E_{l-1}+E_{l+1}),
\end{eqnarray}
where 
\begin{eqnarray}
E_l &=& \frac{1}{4}\int \rho r^2 \left(|a_l|^2 +l(l+1) r^2 |b_l|^2\right)\mathrm{d} r, \\
E_{l-1} &=& \frac{1}{4}\int \rho r^2 l(l-1) r^2 |c_{l-1}|^2\mathrm{d} r, \\
E_{l+1} &=& \frac{1}{4}\int \rho r^2 l(l+2) r^2 |c_{l+1}|^2\mathrm{d} r.
\end{eqnarray}
Note that $\langle Q'_\mathrm{IW}\rangle \propto \Omega^{-2}$, so that relatively rapidly rotating stars are more dissipative. In a given stellar model, we solve Eqs.~\ref{Xeqn}, \ref{PhiEqn} and \ref{WL} numerically using a Chebyshev collocation method with a sufficiently large number of points to ensure accurate solutions. The integrals required to compute Eq.~\ref{QIW} are then straightforward to compute numerically. Note that we fully account for the continuous interior profiles of $\rho(r)$ and $p(r)$; we do not assume a piece-wise homogeneous two-layer model for the star, unlike e.g.~\cite{Mathis2015} and many others.

To compute equilibrium tide dissipation in convection zones, we assume that at each radius in the star, turbulent convection acts like an isotropic effective viscosity $\nu_E(r)$. The latest simulations of \cite{DBJ2020a} suggest that we consider the piece-wise power-law fit 
\begin{eqnarray}
 \nu_{FIT} =  u_{c} l_c \begin{cases}
 5 \quad & (\frac{|\omega|}{\omega_{c}}<10^{-2}), \\
 \frac{1}{2}\left(\frac{\omega_{c}}{|\omega|}\right)^{\frac{1}{2}} \quad &(\frac{|\omega|}{\omega_{c}} \in [10^{-2},5]), \\
\frac{25}{\sqrt{20}}\left(\frac{\omega_{c}}{|\omega|}\right)^2 \quad &(\frac{|\omega|}{\omega_{c}}>5), \\
 \end{cases}
 \label{nuEFIT1}
\end{eqnarray}
where $u_c(r)$ and $l_c(r)$ are the convective velocity and mixing length based on MLT (computed in MESA; see Appendix~\ref{MESA}), $\omega_c(r)=u_c/l_c$ is the convective frequency, and we set $\nu_E=\nu_{FIT}$. Our fit here is based on the maximum value for $\nu_E$ at high frequencies, which provides the maximum dissipation due to this mechanism. This appears to hold for a wide range of Rayleigh numbers in a local model of convection. The resulting dynamic shear viscosity is $\mu(r)=\rho(r)\nu_E(r)$, so that the viscous dissipation is
\begin{eqnarray}
D_\nu = \frac{1}{2} \omega^2\int r^2 \mu(r) D_l(r) \,\mathrm{d} r,
\end{eqnarray}
where the integral is carried out numerically over the entire radial extent of each convection zone, and
\begin{eqnarray}
\nonumber
D_l(r) &=& 3\left|\frac{\mathrm{d}\xi_r}{\mathrm{d} r} -\frac{\Delta_l}{3}\right|^2 +l(l+1)\left|\frac{\xi_r}{r}+r\frac{\mathrm{d}}{\mathrm{d} r}\left(\frac{\xi_h}{r}\right)\right|^2\\
&&\hspace{1.5cm} +(l-1)l(l+1)(l+2)\left|\frac{\xi_h}{r}\right|^2, \\
\Delta_l&=&\frac{1}{r^2}\frac{\mathrm{d}}{\mathrm{d}r}\left(r^2\xi_r\right)-l(l+1)\frac{\xi_h}{r},
\end{eqnarray}
where $\boldsymbol{\xi}_\mathrm{nw}=\xi_r\boldsymbol{e}_r+\boldsymbol{\xi}_h$ is appropriately expanded in terms of $Y_l^m \mathrm{e}^{-\mathrm{i}\omega t}$. The associated tidal quality factor is
\begin{eqnarray}
\label{Qeq}
\frac{1}{Q'_\mathrm{eq}} = \frac{16\pi G}{3(2l+1)R^{2l+1}|A|^2}\frac{D_\nu}{|\omega|}.
\end{eqnarray}

We compute internal gravity wave dissipation in the simplest manner by neglecting rotation (which would otherwise involve more complicated numerical calculations) and by assuming that these waves are launched from the convective/radiative interface and are then fully damped in the radiation zone, following e.g.~\cite{GD1998}. This gives 
\begin{eqnarray}
\label{QIGW}
\frac{1}{Q'_{\mathrm{IGW}}} = \frac{2\left[\Gamma\left(\frac{1}{3}\right)\right]^2}{3^{\frac{1}{3}}(2l+1) (l(l+1))^{\frac{4}{3}}}\frac{R}{GM^2} \mathcal{G} |\omega|^{\frac{8}{3}},
\end{eqnarray}
where the quantities that depend on the radiative/convective interface at $r=r_c$ in a particular stellar model are encapsulated in the quantity 
\begin{eqnarray}
\mathcal{G} = \sigma_c^2 \rho_c r_c^5 \left|\frac{\mathrm{d}N^2}{\mathrm{d}\ln r}\right|_{r=r_c}^{-\frac{1}{3}},
\end{eqnarray}
which takes the value $\mathcal{G}_\odot \approx 2\times 10^{47} \mathrm{kg}\mathrm{m}^2\mathrm{s}^{2/3}$ for the current Sun (in which $\sigma_c=-1.18$), after fitting the local profile of $N^2(r)$.
In this expression, $\rho_c=\rho (r_c)$, and the parameter
\begin{eqnarray}
\sigma_c=\frac{\omega_{\mathrm{dyn}}^2}{A}\left.\frac{\partial \xi_{d,r}}{\partial r}\right|_{r=r_c},
\end{eqnarray}
where the derivative of the dynamical tide radial displacement $\xi_{d,r}$ is determined by integrating the linear differential equation given in Eq.3 in \cite{GD1998}.

\section{MESA Code Parameters}
\label{MESA}

We use MESA version 12778 \citep[e.g.][]{Paxton2011, Paxton2015, Paxton2019}. The inlist file that we use is given below. We alter {\verb|initial_mass|} as required to generate a given stellar model, and the code is stopped manually at a chosen time, usually when the star has left the main sequence.
\begin{verbatim}
&star_job
  create_pre_main_sequence_model = .true.
/ !End of star_job namelist
&controls
! starting specifications
    initial_mass = 1.0
    initial_z = 0.02d0
    MLT_option = 'Henyey'
    max_age = 5.0d10
    max_years_for_timestep = 1.0d8      
    use_dedt_form_of_energy_eqn = .true.
    use_gold_tolerances = .true.
    mesh_delta_coeff = 0.3
    when_to_stop_rtol = 1d-6
    when_to_stop_atol = 1d-6
/ ! end of controls namelist
\end{verbatim}

\end{document}